\documentclass[preprint,preprintnumbers,amsmath,amssymb]{revtex4}
\usepackage[dvips]{graphicx}

\def\M{{\cal M}}

\def\mij{{M_{ij}}}
\def\mtji{{ \left(M^T\right)_{ji} }}

\begin{document}

\title{Accurate image reconstruction from few-views and limited-angle
data in divergent-beam CT}

\author{Emil Y. Sidky}
\email{sidky@uchicago.edu}

\author{Chien-Min Kao}

\author{Xiaochuan Pan}
\email{xpan@uchicago.edu}

\affiliation{%
University of Chicago \\
Department of Radiology \\
5841 S. Maryland Ave., Chicago IL, 60637
}%

\date{\today}% It is always \today, today,

\begin{abstract}
In practical applications of tomographic imaging, there are often
challenges for image reconstruction due to under-sampling and
insufficient data.  In computed tomography (CT), for example, image
reconstruction from few views would enable rapid scanning with
a reduced x-ray dose delivered to the patient. Limited-angle
problems are also of practical significance in CT.  In this work,
we develop and investigate an iterative image reconstruction
algorithm based on the minimization
of the image total variation (TV) that applies to divergent-beam CT.
Numerical demonstrations of our TV algorithm are performed with
various insufficient data problems in fan-beam CT.
The TV algorithm can be generalized to cone-beam CT as well
as other tomographic imaging modalities.
\end{abstract}

%\pacs{Don't forget me}

\maketitle

\section{Introduction}
\label{sec:intro}

In various forms of tomography, one of the main issues for image
reconstruction centers on data sufficiency and on how to estimate a
tomographic image when the projection data are not theoretically
sufficient for exact image reconstruction.  Insufficient data problems
occur quite frequently because
of practical constraints due to the imaging hardware, scanning geometry,
or ionizing radiation exposure.  The insufficient data problem can take
many forms, but the forms that we shall consider in this work relate
to divergent-beam x-ray computed tomography (CT).  One aspect of the insufficient
data problem derives from sparse samples; namely, we will consider
image reconstruction from projection data at few views. 
We will also consider
two other imperfect scanning data situations: limited angular range and gaps
in the projection data caused by bad detector bins. In each of these
three examples, the projection data are not sufficient
for exact reconstruction of tomographic images and application of standard
analytic algorithms such as filtered back-projection (FBP) will lead
to conspicuous artifacts in reconstructed images.

There have been a number of algorithms proposed to overcome data insufficiency
in tomographic imaging, and there are essentially two types of approaches.
First, one can interpolate or extrapolate the missing data regions from
the measured data set, followed by analytic reconstruction.
Such approaches may be useful for a specific scanning
configuration, imaging a particular object. It is, however, difficult to make
general conclusions on the utility of such an approach.
Second, one can employ
an iterative algorithm to solve the data model for images
from the available measurements.
Numerous iterative
algorithms have been used for tomographic image reconstruction with
varying degrees of success. These algorithms differ 
in the constraints that
they impose on the image function, the cost function that they seek
to minimize, and the actual implementation of the iterative scheme.
This article follows the second approach.

Two widely used iterative algorithms for tomographic imaging are the
algebraic reconstruction technique (ART) (see e.g.
Chap. 11 in Ref. \cite{Herman:80}
and Sec. 5.3.1 in Ref. \cite{Natterer:01}) and the
expectation-maximization (EM) algorithm (see e.g.
Sec. 5.3.2 in Ref. \cite{Natterer:01}
and Sec. 15.4.6 in Ref. \cite{Barrett:04}).   For the case
where the data are consistent yet are not sufficient to determine a unique
solution to the imaging model,
the ART algorithm
finds the image that is consistent with the data and minimizes
the sum-of-squares of the image pixel values.  The EM
algorithm applies to positive
integral equations, which is appropriate for the CT-imaging model, and seeks
to minimize the Kullback-Liebler distance between the measured data and
the projection of the estimated image.
%Both of these algorithms
%make different assumptions about the image function, and as a result yield
%different results.   One of the factors that determines the utility
%of each method is how appropriate the assumption made on the image function
%is.  For example, the MLEM method requires that the image be non-negative,
%a valid assumption for CT attenuation coefficients.
Part of the success of the EM algorithm
derives from the fact that the positivity constraint is built in
to the algorithm, and that it is relatively
robust against data inconsistencies introduced
by signal noise.

For certain imaging problems, an accurate iterative scheme can be
derived for the imperfect sampling problem by making a strong assumption
on the image function.  For example, in the reconstruction
of blood vessels from few-view projections,
one can assume that the 3D blood-vessel structure is sparse.
It is possible to design an effective iterative algorithm that seeks
a solution from sparse projection data.  This can be accomplished by minimizing
the $\ell_1$-norm of the image constrained by the fact that the image yields
the measured projection data \cite{Kudo:02}.
The $\ell_1$-norm of the image is simply the
sum of the absolute values of the image pixel values, and its minimization
subject to linear constraints
leads to sparse solutions \cite{Kudo:02,Daubechies:04}.

% To see that this
% scheme favors sparse solutions, we borrow the example from Ref. \cite{Kudo:02}.

% For a linear system with two unknowns, say $x_1$ and $x_2$,
% it is well-known that a unique
% solution is specified by two independent linear equations.  Suppose instead
% that we have only a single linear equation, as illustrated in Fig. \ref{l1explanation},
% and we know that the solution is sparse (either $x_1$ or $x_2$ is non-zero).
% Because of the shape of the $ll_1$-ball,
% a sparse solution is selected by finding a point on the line with minimal
% $\ell_1$-norm.   Note that minimizing the $\ell_2$-norm, which corresponds
% to ART, yields a non-sparse solution.  This example is a bit oversimplified
% compared to a full-blown tomographic imaging problem, but we use it just
% to illustrate how $\ell_1$ minimization yields sparse solutions and that different
% objective functions can lead to very different solutions particularly for
% missing data problems.  The results shown in Ref. \cite{Kudo:02} for the
% 3D blood vessel reconstructions were nothing short of amazing given that
% very few projection views were employed in the reconstructions.  

%Though impressive, the results of Ref. \cite{Kudo:02} only apply to image
%functions that are sparse.
In medical and other tomographic imaging applications, images
are generally extended distributions, violating the prerequisite of employing
the $\ell_1$-based algorithms.  There is, however, a similar sparseness property
that does describe a wide class of tomographic images.  Often times in medical
and other applications, tomographic images are relatively
constant over extended volumes, for
example within an organ. Rapid variation in the image may only occur at boundaries of internal
structures.  Thus an image itself might not be sparse, but the image formed
by taking the magnitude of its gradient could be approximately
sparse \cite{candes-robust}.

\begin{figure}[ht]

\begin{minipage}[b]{0.48\linewidth}
\centering
\centerline{\includegraphics[width=6cm,clip=TRUE]{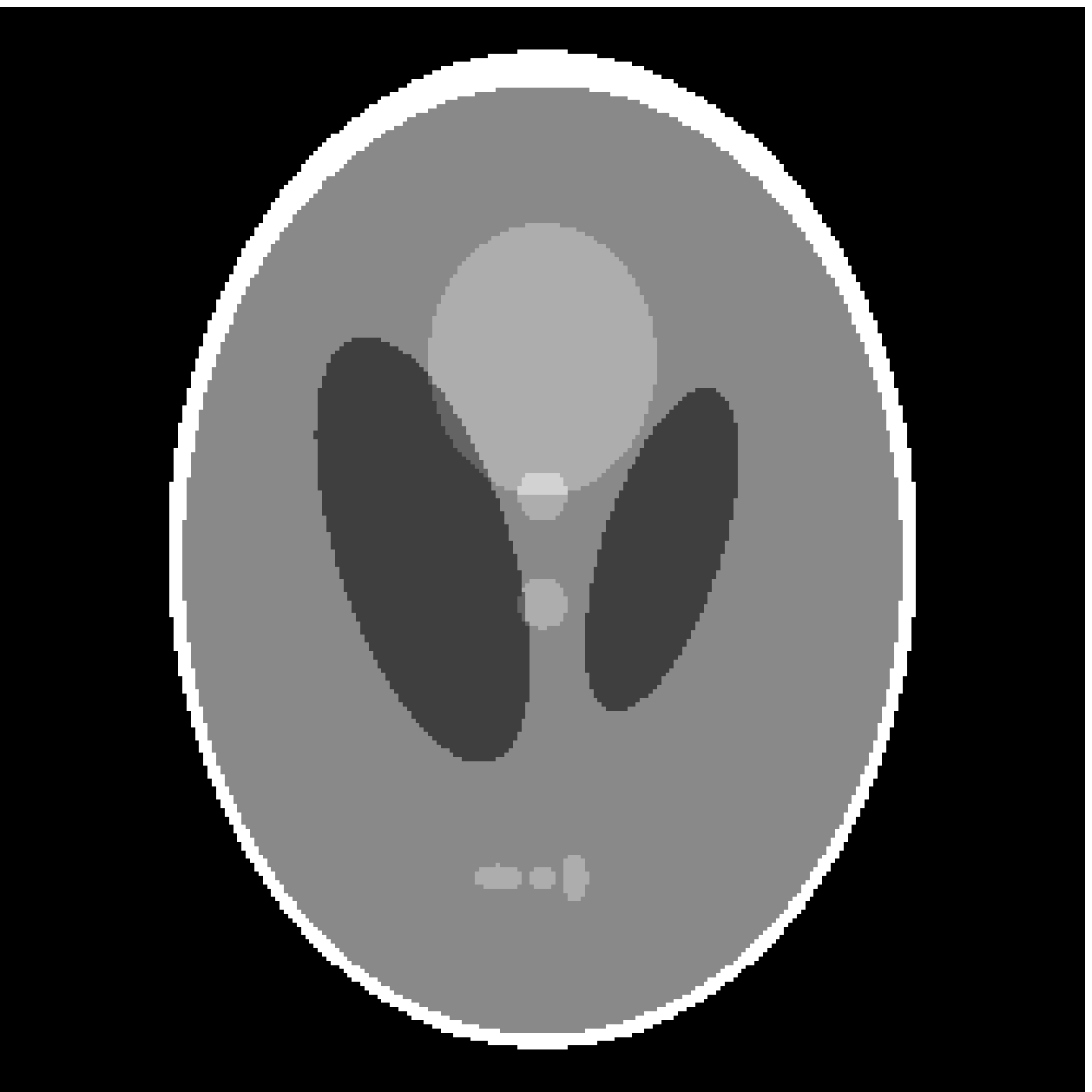}}
\end{minipage}
\begin{minipage}[b]{0.48\linewidth}
\centering
\centerline{\includegraphics[width=6cm,clip=TRUE]{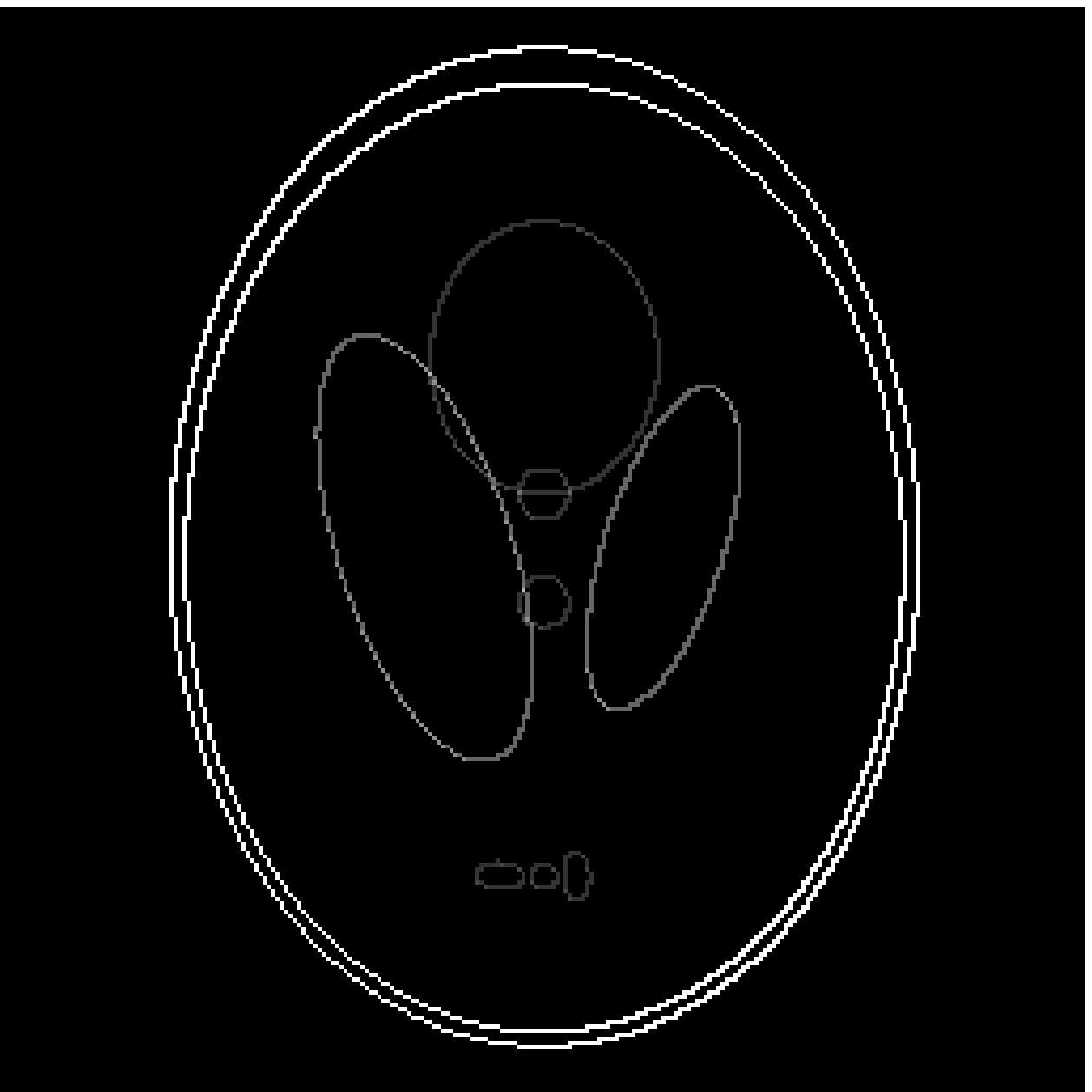}}
\end{minipage}
\caption{
Left: Shepp-Logan phantom shown in a gray scale window of [0.87,1.15].
Right: Magnitude of the image gradient of the Shepp-Logan phantom.
Note the sparseness of the gradient image.
\label{fig:shepp}}
\end{figure}

We demonstrate
this point with the widely used Shepp-Logan phantom in Fig. \ref{fig:shepp}.
If the pixel values are labeled by $f_{s,t}$, the image gradient magnitude
is:
\begin{equation}
\label{grad_mag}
|\vec{\nabla}f_{s,t}|=\sqrt{(f_{s,t}-f_{s-1,t})^2 + (f_{s,t}-f_{s,t-1})^2}.
\end{equation}
We refer to this quantity as the gradient image.  
The number
of non-zero pixels in this 256x256 image is 32,668 , while the number
of non-zero pixels in its gradient image is  only 2,183. 

To develop an iterative
algorithm that takes advantage of this sparseness,
the objective function to be minimized
is the $\ell_1$-norm of the gradient image, otherwise known as the total variation
(TV) of the image:
\begin{equation}
\label{totalVar}
||f_{s,t}||_{TV}= \sum_{s,t} |\vec{\nabla}f_{s,t}| =
\sum_{s,t} \sqrt{(f_{s,t}-f_{s-1,t})^2 + (f_{s,t}-f_{s,t-1})^2}.
\end{equation}
TV has been utilized in image
processing for denoising of images while preserving edges
\cite{Rudin:92,Vogel:96}, and TV has been suggested as
a regularization function in Bayesian reconstruction,
implemented in an EM algorithm \cite{Panin:99,PerssonSPECT:01,PerssonLimited:01}.
The use of the image TV here is different in that we seek an algorithm
that is an implementation of an optimization program,
which yields possibly the {\it exact} image for sparse data problems
under the condition of exact data consistency.

Recently, a TV-based algorithm for recovering an image from sparse
samples of its Fourier transform (FT) 
was developed \cite{candes-robust,candes-ERP}.
In that work the authors investigated the optimization program of
minimizing the image TV under the constraint that the FT of the
image matches the known FT samples. They showed that this optimization
program satisfies an ``exact reconstruction principle'' (ERP) for
sparse data: if the number
of FT samples is twice the number of non-zero pixels in the
gradient image, then this optimization program
can yield a unique solution, which is in fact the true image for
almost every image function.
The algorithm for FT inversion from sparse samples was
applied to image reconstruction from 2D parallel-beam data
at few-views. The use of the FT-domain
TV algorithm (FT-TV) to address the 2D parallel-beam
problem was possible because of the
central slice theorem, which links the problem
to FT inversion.

The FT-TV algorithm, however, cannot
be applied to image
reconstruction for divergent-beam CT, i.e. fan-beam and cone-beam CT,
because there is no central slice theorem to bring the projection
data into the image's Fourier space.  The ERP
of the FT-TV algorithm may extend to inversion of other linear systems
from sparse data \cite{candes-ERP}.
To our knowledge, no TV-based algorithm that exploits the
ERP and that is specific to tomographic image
reconstruction from divergent-beam projections has been developed
previously.

In this work, we investigate and develop a TV algorithm for
image reconstruction from divergent-beam projections applicable to both
fan-beam and cone-beam CT imaging. We present a TV
iterative algorithm that can reconstruct accurate
images from sparse or insufficient
data problems that may occur due to practical issues of CT
scanning. The sparse data problem that we consider here
is reconstruction from few-view projections; whereas the insufficient data
problems that we investigate are reconstructions from data acquired over
a limited angular range
or with a detector containing gaps due to bad detector bins.
Much research has been done on image reconstruction algorithms
for the few-view problem, see e.g. Refs. \cite{Kudo:02,Wu:03,KoleI:03},
and for the limited angular range problem,
see e.g. Refs. \cite{PerssonLimited:01,Bresler:98,Rantala:06}.
Comparison with these algorithms is a topic for future work,
but we do present comparisons with the basic EM and ART algorithms.
We point out that the comparison with EM and ART are only meant
to reveal the ill-posedness of the imaging problems considered here.
In Sec. \ref{sec:theory}, we describe our TV iterative algorithm
for tomographic image reconstruction from divergent-beam projection data.
In Sec. \ref{sec:results},
we demonstrate and validate the proposed TV algorithm for image
reconstruction in
various sparse or insufficient
data problems.  In Sec. \ref{sec:compfac}, we re-examine some of the examples
of Sec. \ref{sec:results} under non-ideal conditions such as data
inconsistency due to noise. Although the numerical results involve
only fan-beam CT, the same algorithm can readily
be applied to image reconstruction
in cone-beam CT.

\section{Method}
\label{sec:theory}

In this section, we describe our TV algorithm
for image reconstruction in divergent-beam CT.  The image function is
represented in its discrete form as a vector 
$\vec{f}$ of length $N_\text{image}$ with
individual elements $f_j$, $j=1,2,\dots,N_\text{image}$.
When it is necessary to
refer to pixels in the context of a 2D image we use 
the double subscript form $f_{s,t}$, where
\begin{equation}
\label{indexConv}
j=(s-1) W +t; \; \; \; \; \; s=1,2,\dots,H; \; \; \; \;
t=1,2,\dots,W;
\end{equation}
and integers $W$ and $H$ are, respectively,
the width and height of the 2D image array, which has a total number
of pixels $N_\text{image}=W \times H$.
The projection-data vector $\vec{g}$ has 
length $N_\text{data}$ with individual
measurements referred to as $g_i$, $i=1,2,\dots,N_\text{data}$.

The general theoretical setting for the TV algorithm discussed here involves
inversion of a discrete-to-discrete linear transform,
\begin{equation}
\label{linearSystem}
\vec{g}=\M \vec{f},
\end{equation}
where the system matrix $\M$ is composed of $N_\text{data}$
row vectors $\vec{M}_i$ that yield each data point, $g_i=\vec{M}_i \cdot
\vec{f}$. The individual elements of the system matrix are $M_{ij}$.
We seek to
obtain an image represented by the finite vector $\vec{f}$
from knowledge of
the data vector $\vec{g}$
and the system matrix $\M$.
Mathematically, the problems we consider here
involve insufficient data;
namely the number of data samples $N_\text{data}$ is
not enough to uniquely determine the $N_\text{image}$
values of the image vector $\vec{f}$  by directly inverting
Eq. (\ref{linearSystem}).
The overall strategy is to incorporate the assumption of
gradient image sparseness on the image
function $\vec{f}$ to arrive at a solution from knowledge of
the data $\vec{g}$.

To solve the linear system represented in Eq. (\ref{linearSystem})
we develop a TV algorithm that implements the following 
optimization program \cite{candes-stable}: Find $\vec{f}$ that
\begin{equation}
\label{optimization}
\min \| \vec{f} \|_{TV} \; \; \; \text{such that} \; \; \;
\M \vec{f} = \vec{g}, \; \; \; f_j \ge 0.
\end{equation}
In the algorithm,
the minimization of the image TV is performed by the
gradient descent method,
and the constraints imposed by the known projection data 
are incorporated by projection on convex sets (POCS)
(see e.g. Sec. 15.4.5 of Ref. \cite{Barrett:04}).
We use POCS for enforcing the projection data constraint, because,
even in the case of sparse sampling, the size of the projection data sets
can be large, and POCS can efficiently handle large data sets.
In the following we define the system matrix used for modeling
the divergent-beam projections, and describe the TV algorithm for
implementing the program in Eq. (\ref{optimization}).  We conjecture
that the linear system matrices corresponding to the various
scanning configurations studied in Secs. \ref{sec:results}
and \ref{sec:compfac}, below, support an ERP
for insufficient data, and
we demonstrate this possibility with numerical examples.

\subsection{System matrix for the divergent-beam configuration}

\begin{figure}[ht]

\begin{minipage}[b]{0.98\linewidth}
\centering
\centerline{\includegraphics[width=8cm,clip=TRUE]{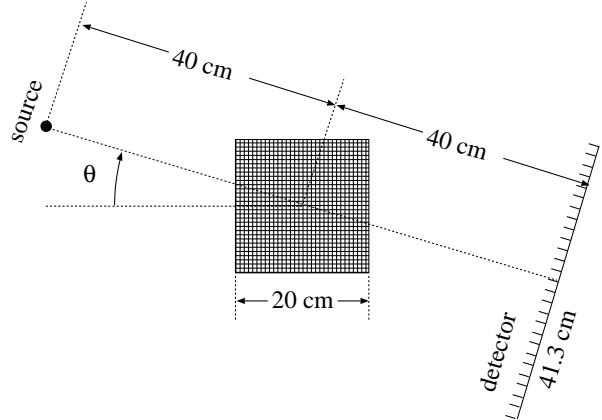}}
\end{minipage}
\caption{
Fan-beam CT configuration used in the simulations presented in this
work.  The 256x256 image array is 20x20 cm$^2$, and the detector
is composed of 512 bins.  The center of rotation for the source-detector
gantry coincides with the center of the image array.
\label{fig:fanConfig}}
\end{figure}

\begin{figure}[ht]

\begin{minipage}[b]{0.98\linewidth}
\centering
\centerline{\includegraphics[width=8cm,clip=TRUE]{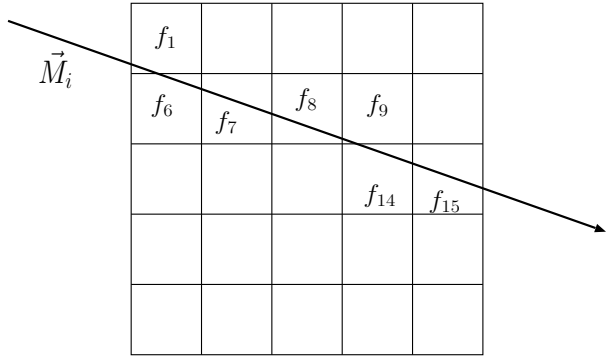}}
\end{minipage}
\caption{
The ray passing through this 5x5 image array illustrates an individual
row vector of the system matrix $\vec{M}_i$.  In this case the data
point $d_i$ is calculated as $d_i=\sum_{j=1}^{25} M_{ij}f_j$, where $M_{ij}$
is the length of the $i$th ray traversing the $j$th pixel. The illustrated
system matrix has non-zero entries only on image pixels
$f_1$, $f_6$, $f_7$, $f_8$, $f_9$, $f_{14}$, and $f_{15}$ .
\label{fig:sysMat}}
\end{figure}

In divergent-beam CT, the x-ray source is a single spot for each
projection view.  The projection data are captured on a 1D or
2D detector
array for the fan-beam or cone-beam system.
For illustration and introduction of the configuration used in the
Secs. \ref{sec:results} and \ref{sec:compfac},
we focus on the fan-beam configuration shown in
Fig. \ref{fig:fanConfig}. The detector is modeled as a straight-line
array of 512 detector bins, which is large enough so that
the field-of-view is the circle inscribed
in the 256x256 imaging array.
The CT measurements can be
related to the path integral of the x-ray attenuation coefficient
along the rays defined by the source spot and individual detector bins.
In the discrete setting, these ray integrals can be written as
weighted sums over the pixels traversed by the source-bin ray as
\begin{equation}
\label{sysMat}
d_i = \sum_{j=1}^{N_\text{image}} M_{ij} f_j , \; \; \; \; \; 
\text{where} \; \; \; \; \;
i=1,2,\dots,N_\text{data}.
\end{equation}
To model the fan-beam projection of the discrete image array,
we employ the ray-driven projection model where the system
matrix weights $M_{ij}$ are computed by calculating the intersection
length of the $i$th ray through the $j$th pixel. The ray-driven
system matrix is illustrated for a 5x5 image array
in Fig. \ref{fig:sysMat}.  There are other ways to model the
discrete projection such as pixel-driven and distance-driven
models \cite{Basu:04}, which provide alternative definitions of pixel
weights.
Even though the system matrix
discussed here is for the fan-beam configuration, extension to
cone-beam 3D imaging is straight-forward.

An interesting difference, between the TV algorithm presented here and
the FT-TV algorithm of Ref. \cite{candes-robust}, is the construction of
the system matrix.  In Ref. \cite{candes-robust}, the 2D parallel-beam data
are processed by taking a 1D FT along the detector coordinate, and the system
matrix is the discrete 2D FT.  In this work the system matrix represents
directly the discrete ray integration of the image, and there is no transformation
of the projection data.  Thus, even in the limit that the focal length of the
fan-beam tends to infinity, our TV algorithm
does not yield the FT-TV algorithm.

\subsection{Computation of TV gradient and realization of data constraint}

The TV algorithm minimizes
the TV of the image estimate, which can be accomplished by use of
the gradient descent method \cite{candes-robust} and
other optimization methods. Performing the
gradient descent requires the
expression for the gradient of the image TV.
This gradient can also be thought of as an image, where each pixel
value is the partial derivative of the image TV with
respect to that pixel.  Taking the derivative of $\left\| \vec{f} \right\|_{TV}$
with respect to each pixel value results in a singular expression.
We thus use the following approximate derivative,
\begin{multline}
\label{tvgrad}
v_{s,t}=
\frac{ \partial \left\| \vec{f} \right\|_{TV} }
{ \partial f_{s,t}} \approx
\frac{2 \left(f_{s,t} - f_{s-1,t} \right) +
2 \left(f_{s,t} - f_{s,t-1} \right)}
{\sqrt{ \epsilon +
\left(f_{s,t} - f_{s-1,t}\right)^2 +
\left(f_{s,t} - f_{s,t-1}\right)^2 }} \\
-
\frac{2 \left(f_{s+1,t} - f_{s,t} \right)}
{\sqrt{ \epsilon +
\left(f_{s+1,t} - f_{s,t}\right)^2+
\left(f_{s+1,t} - f_{s+1,t-1}\right)^2 }}  \\
- \frac{2 \left(f_{s,t+1} - f_{s,t} \right)}
{\sqrt{ \epsilon +
\left(f_{s,t+1} - f_{s,t}\right)^2+
\left(f_{s,t+1} - f_{s-1,t+1}\right)^2 }},
\end{multline}
where $\epsilon$ is a small positive number; for the
results below we used $\epsilon= 10 ^{-8}$.
Note that this expression is valid for non-border pixels.
We refer to
the resulting gradient vector as $\vec{v}$, and just as with
the image vector, its individual elements can be denoted by either
a single index $v_j$ or pixel indexes $v_{s,t}$. In the actual
algorithm, we employ the normalized TV
gradient $\hat{v}$.  

We use the POCS method to realize the linear system constraints
in Eq. (\ref{optimization}).
Each measured point $g_i$ of the data vector specifies a hyperplane
in the $N_\text{image}$-dimensional space of all possible solutions $\vec{f}$. 
The basic POCS method projects the current estimate of $\vec{f}$
onto the hyperplanes, which are convex sets,
corresponding to each data point in sequential order.
By repeating this process the image estimate moves toward
the intersection of all of these hyperplanes, 
which is the sub-space of valid solutions to the linear system.
In our POCS implementation,
we will also include the positivity constraint.

\subsection{TV algorithm for divergent-beam CT}
\label{sec:tvAlgorithm}

Having specified the system matrix,
TV gradient, and data constraints,
we now describe the iterative steps of
the TV algorithm, which implements
the optimization program described in Eq. (\ref{optimization})
for image reconstruction from divergent-beam data.
Each iteration 
consists of two phases: POCS and gradient descent. The
POCS phase is further broken down into two steps that enforce
data consistency and positivity. As a result, the steps
comprising each loop are:
the DATA-step, which
enforces consistency
with the projection data;  the POS-step, which ensures
a non-negative image; and  the GRAD-step, which reduces the TV
of the image estimate. The iteration performed
in the algorithm has two levels: the overall iteration number
is labeled by $n$, and the sub-iterations in the DATA- and GRAD-steps
are labeled by $m$. The image vector during the
iterations of the DATA-step
is $\vec{f}^{(TV-DATA)}[n,m]$, indicating the $m$th DATA-step
sub-iteration within the $n$th iteration. We use $\vec{f}^{(TV-POS)}[n]$
to denote the image estimate after projection onto the non-negative
half-plane. Finally, $\vec{f}^{(TV-GRAD)}[n,m]$ represents the $m$th
gradient descent step within the $n$th iteration. The steps
of the algorithm are: \\
(A) Initialization:
\begin{equation}
\label{initialization}
n=1 \; \; \; \text{and} \; \; \;
\vec{f}^{(TV-DATA)}[n,1]=0;
\end{equation}
(B) Data projection iteration, for $m=2,\dots,N_\text{data}$:
\begin{equation}
\label{artstep}
\vec{f}^{(TV-DATA)}[n,m]=\vec{f}^{(TV-DATA)}[n,m-1]- \vec{M}_{m-1}
\frac{g_{m-1}-\vec{M}_{m-1} \cdot \vec{f}^{(TV-DATA)}[n,m-1]}
{\vec{M}_{m-1} \cdot \vec{M}_{m-1}};
\end{equation}
(C) Positivity constraint:
\begin{equation}
\label{positivity}
(f_j)^{(TV-POS)}[n]=
\begin{cases}
(f_j)^{(TV-DATA)}[n,N_\text{data}] & (f_j)^{(TV-DATA)}[n,N_\text{data}] \ge 0 \\
0 & (f_j)^{(TV-DATA)}[n,N_\text{data}] < 0
\end{cases};
\end{equation}
(D) TV gradient descent initialization:
\begin{gather}
\label{gradinit}
\vec{f}^{(TV-GRAD)}[n,1]=\vec{f}^{(TV-POS)}[n]; \notag \\
d_A(n)=\left\| \vec{f}^{(TV-DATA)}[n,1] - \vec{f}^{(TV-POS)}[n]\right\|_2;
\end{gather}
(D$^\prime$) TV gradient descent, for $ m=2,\dots,N_\text{grad}$:
\begin{gather}
\vec{v}_{s,t}[n,m-1]=
\left. \frac{\partial \| \vec{f} \|_{TV}} {\partial f_{s,t}}
\right|_{f_{s,t}=f^{(TV-GRAD)}_{s,t}[n,m-1] }; \; \; \; \; \;
\hat{v}[n,m-1]= \frac{\vec{v}[n,m-1]}{\left| \vec{v}[n,m-1] \right|} ; \notag \\
\vec{f}^{(TV-GRAD)}[n,m] =\vec{f}^{(TV-GRAD)}[n,m-1] - a {d_A(n)}
\hat{v}[n,m-1];
\label{gradstep}
\end{gather}
(E) Initialize next loop:
\begin{equation}
\label{reinit}
\vec{f}^{(TV-DATA)}[n+1,1]=\vec{f}^{(TV-GRAD)}[n,N_\text{grad}];
\end{equation}
Increment $n$ and return to step (B).
In the text, when we refer to the iteration number of the TV algorithm,
we mean the
iteration number of the outer loop indicated by the index $n$.
The iteration is stopped
when there is no appreciable change in the intermediate images
after the POCS steps; namely the difference between
$\vec{f}^{(TV-POS)}[n]$ and $\vec{f}^{(TV-POS)}[n-1]$ is ``small''.

The distance $d_A(n)$ provides a measure for the difference
between the image estimate before the DATA-step and the
estimate after the enforcement of positivity.  The gradient descent
procedure is controlled by specifying the parameter
$a$, the fraction of the distance $d_A(n)$
along which the image is incremented, and $N_{grad}$ the total number of
gradient descent steps that are performed.  The algorithm relies
on the balance between the POCS phase (DATA- and POS-steps) and the
gradient descent.
By scaling the size of the gradient descent
step with $d_A(n)$, the relative importance of the POCS and
gradient descent stages of
the algorithm maintains this balance.  
As long as the total change in the image
due to the gradient descent does not exceed the change in the image
due to POCS the overall iteration steps will steer the image estimates
closer to the
solution space of the imaging linear system.

If the step size of
the gradient descent is too strong the image becomes uniform and
inconsistent with the projection data. On the other hand,
if the step size of the gradient descent
is too small, the algorithm reduces to standard ART with a positivity
constraint included.
For the results shown in this article we selected
$a=0.2$, and $N_{grad}=20$. 
These values strikes a good balance between the POCS
steps and the TV-gradient descent, and they seem to work well for
a wide range of reconstruction problems, including those
addressed in Secs.
\ref{sec:results} and \ref{sec:compfac} below.  The algorithm
appears to be robust in that changes to the parameters only appear
to alter the convergence rate and not
the final image. Further investigation of the algorithm
parameters may improve the convergence speed.

\section{Numerical Results: ideal conditions}
\label{sec:results}

For the results in this section, we demonstrate and validate our TV
algorithm under ``ideal'' conditions.  The true image solution
is taken to be the Shepp-Logan image shown in Fig. \ref{fig:shepp} discretized
on a 256x256 pixel grid.  This phantom is often used in evaluating
tomographic reconstruction algorithms. As also shown in Fig. \ref{fig:shepp},
its gradient image is sparse with only 2,183 non-zero pixels.
This number is roughly only 6.7\% of the 
32,668 non-zero pixels of the Shepp-Logan image itself.
Taking the result for Fourier inversion \cite{candes-robust}
as a rule of thumb for the current problem,
one might expect that a minimum of twice as many non-zero, independent
projection measurements are needed for obtaining the image. Thus we suppose
that a minimum of 4,366 measurements are required for the ERP.
We first demonstrate the image recovery from sparse data with the few-view
example shown below.  Subsequently, we show
the utility of the TV algorithm for
other insufficient data
problems where there are plenty of projection ray measurements,
but the angular or projection coverage is less than the minimum for
analytic reconstruction.
The insufficient data problems demonstrated below are the limited scanning angle
problem, and the ``bad bins'' problem where there is a gap on the detector
for all available projection views.

For the numerical experiments presented here,
the simulated fan-beam
configuration are variations on the
configuration shown in Fig. \ref{fig:fanConfig}.
In the first set of experiments
the data used are ideal in the sense that they
are the exact line integrals,
up to round-off error in the computer, of the
discrete 256x256 Shepp-Logan image.  The data are, however,
severely under-determined
so that there would be no chance
of directly solving the linear equation,
in Eq. (\ref{linearSystem}).
The detector modeled has 512 bins, and the total number of measured rays is 512
multiplied by the number of view angles.  But the important number is actually
the total number of non-zero measurements, and this is stated with each example
discussed below.

In order to illustrate the degree of ill-posedness for each numerical example,
we compare the proposed TV algorithm with standard EM and ART
algorithms,
which have been widely applied to solving the 
under-determined or unstable linear systems in 
tomographic imaging.
A unique feature of
EM is that the positivity constraint is built-in to the algorithm, and
for CT imaging applications the object function is
positive.
The EM implementation used here is basic,
specified by the following update equation:
\begin{equation}
\label{EMupdate}
f_j^\text{(EM)}[n]=f_j^\text{(EM)}[n-1]
\frac{\sum_i \mtji \frac{g_i}{\sum_j \mij f_j^\text{(EM)}[n-1]}}
{\sum_i \mtji}.
\end{equation}
In our studies, we used no regularization during the iterations.

The ART algorithm is identical to the TV algorithm discussed in
Sec. \ref{sec:tvAlgorithm} except that step (D), the minimization of
the image TV, is not performed.  
The steps for the ART algorithm are: \\
(A) Initialization
\begin{equation}
\label{initialization-ART}
n=1; \; \; \; \; \;
\vec{f}^{(ART-DATA)}[n,1]=0;
\end{equation}
(B) Data-projection iteration, for $m=2,\dots,N_\text{data}$:
\begin{equation}
\label{artstep-ART}
\vec{f}^{(ART-DATA)}[n,m]=\vec{f}^{(ART-DATA)}[n,m-1]- \vec{M}_i 
\frac{g_i-\vec{M}_i \cdot \vec{f}^{(ART-DATA)}[n,m-1]}
{\vec{M}_i \cdot \vec{M}_i};
\end{equation}
(C) Positivity constraint:
\begin{equation}
\label{positivity-ART}
(f_j)^{(ART-POS)}[n]=
\begin{cases}
(f_j)^{(ART-DATA)}[n,N_\text{data}] & (f_j)^{(ART-DATA)}[n,N_\text{data}] \ge 0 \\
0 & (f_j)^{(ART-DATA)}[n,N_\text{data}] < 0
\end{cases};
\end{equation}
(D) Initialization next loop:
\begin{equation}
\label{reinit-ART}
\vec{f}^{(ART-DATA)}[n+1,1]=\vec{f}^{(ART-POS)}[n];
\end{equation}
Increment $n$ and return to step (B). Again, no explicit
regularization was performed during the ART iterations.
For both EM and ART algorithms
the iteration was stopped when there was
no appreciable change in the image.

No explicit regularization for the EM and ART algorithms was 
used for two reasons. First, we wish to demonstrate only the
degree of ill-posedness of the linear systems corresponding
to the various scanning configurations investigated below. And
this is effectively demonstrated by using well-known algorithms
such as EM and ART.  Second, we are comparing the TV algorithm
with the EM and
ART algorithms
on how well they solve the linear system corresponding to
sparse sampling or insufficient projection data.
The data used
for the bulk of the examples are ideal (up to machine precision), and
any explicit regularization during the EM or ART
iterations would introduce inconsistency
between the reconstructed image and the projection data.

\subsection{Few-view results}
\label{sec:fewView}

\begin{figure}[ht]

\begin{minipage}[b]{0.8\linewidth}
\centering
\centerline{\includegraphics[width=15cm,clip=TRUE]{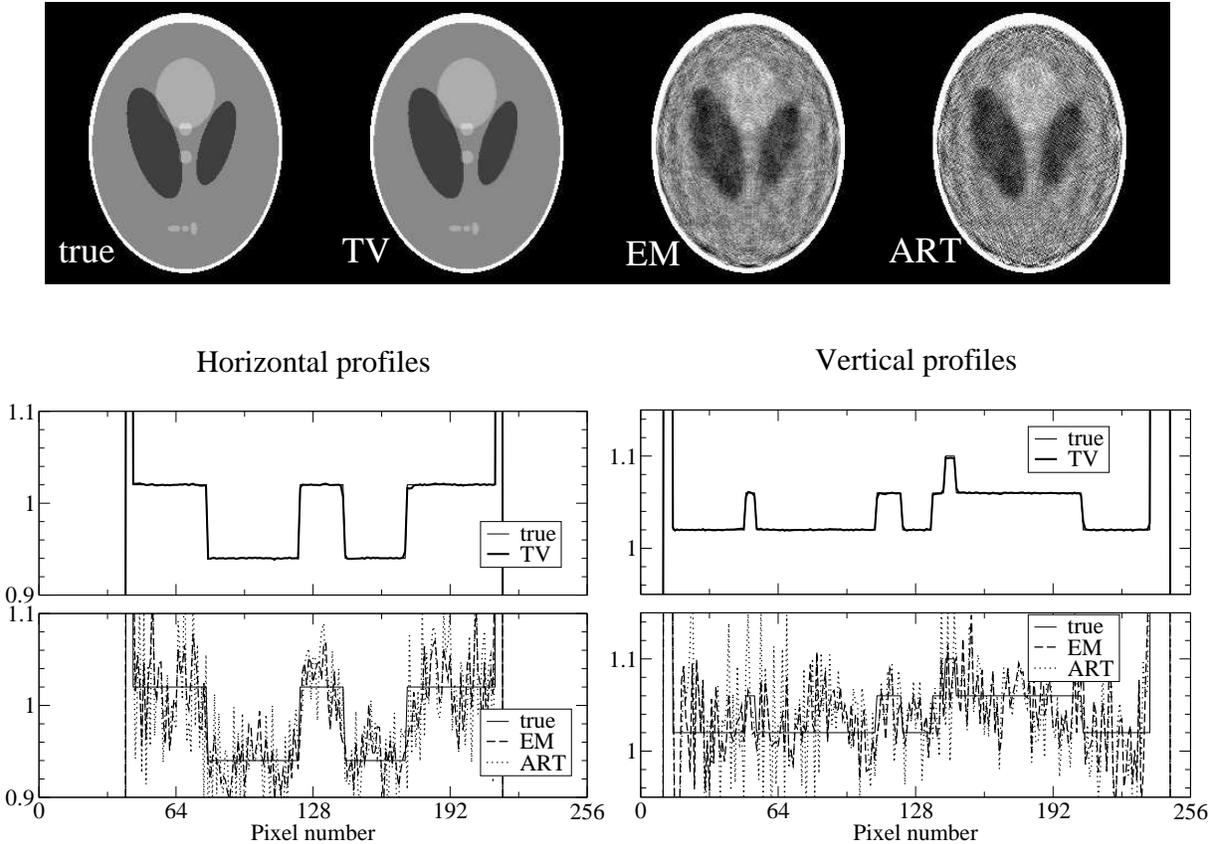}}
\hskip 1.0cm
\end{minipage}
\begin{minipage}[b]{0.48\linewidth}
\centering
\centerline{\includegraphics[width=8cm,clip=TRUE]{figs/fewViewX.eps}}
\end{minipage}
\begin{minipage}[b]{0.48\linewidth}
\centering
\centerline{\includegraphics[width=8cm,clip=TRUE]{figs/fewViewY.eps}}
\end{minipage}
\caption{
Upper row: The true image and images reconstructed by use of the 
TV, EM, and ART algorithms
from 20-view projection data. The display gray scale is [0.85,1.15]. 
Middle row: Image profiles along the centers
of the images in the horizontal  and 
vertical directions obtained with the TV algorithm (thick line).  
Lower row: Image profiles along the centers
of the images in the horizontal and 
vertical directions obtained with the EM
(dashed lines) and ART (dotted lines) algorithms.  
The corresponding true profiles are plotted as the
thin lines in the middle and lower rows. 
\label{fig:fewView}}
\end{figure}

The first case is a reconstruction problem from 
few-view projections in fan-beam CT.  Using the
Shepp-Logan phantom shown in Fig. \ref{fig:fewView}A,
we generated projection data at 20 view angles specified by:
\begin{equation}
\label{fewViewAngles}
\theta_i=
\begin{cases}
18^\circ * (i-1) & 1 \le i \le 10 \\
18^\circ * (i-0.5) & 10 < i \le 20.
\end{cases}
\end{equation}
Though sparse, the angles cover 360$^\circ$ about the object. The shift
in the second half of the angular measurements helps to reduce redundancy
in the scanned data. The total number of measurement rays is 
$512 \times 20 =10,240$, 
but only 8,236 of these projection elements are non-zero.  
This number is larger than the twice the support of 
the gradient image, but it
is well below the support of the Shepp-Logan phantom itself.  In addition,
the angular direction  is severely undersampled.

From the projection data generated at the 20 views, 
we reconstructed images, as shown in row one of Fig. \ref{fig:fewView},
by use of the TV, EM, and ART algorithms.
The number of iterations for each algorithm was 200.
For a quantitative comparison, we also compare the image profiles
along the central lines of the images in 
the horizontal and vertical directions.
The results in Fig. \ref{fig:fewView} indicate that 
the TV reconstruction is visually indistinguishable from the true
image, 
suggesting that the system matrix
corresponding to sparse fan-beam data may have the ERP
even though the column vectors of the system
matrix do not form an ortho-normal basis. 
The EM and ART results show considerable artifacts.

\begin{figure}[ht]

\begin{minipage}[b]{0.8\linewidth}
\centering
\centerline{\includegraphics[width=12cm,clip=TRUE]{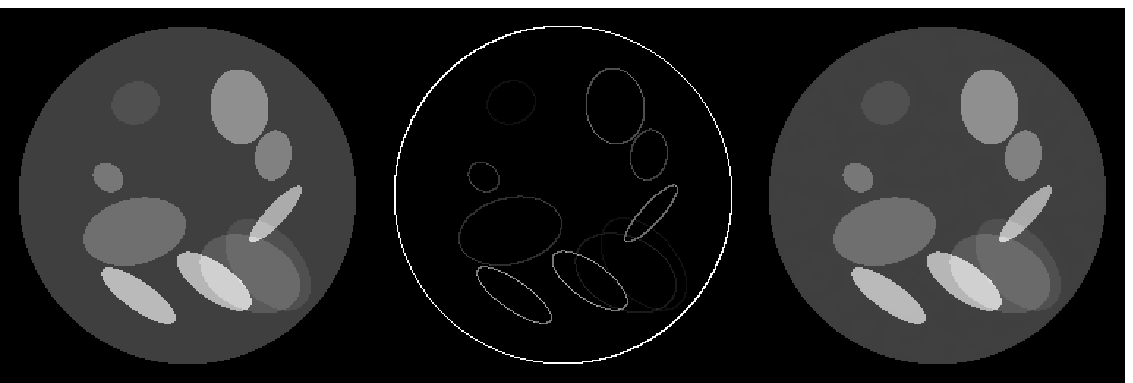}}
\end{minipage}
\begin{minipage}[b]{0.8\linewidth}
\centering
\centerline{\includegraphics[width=12cm,clip=TRUE]{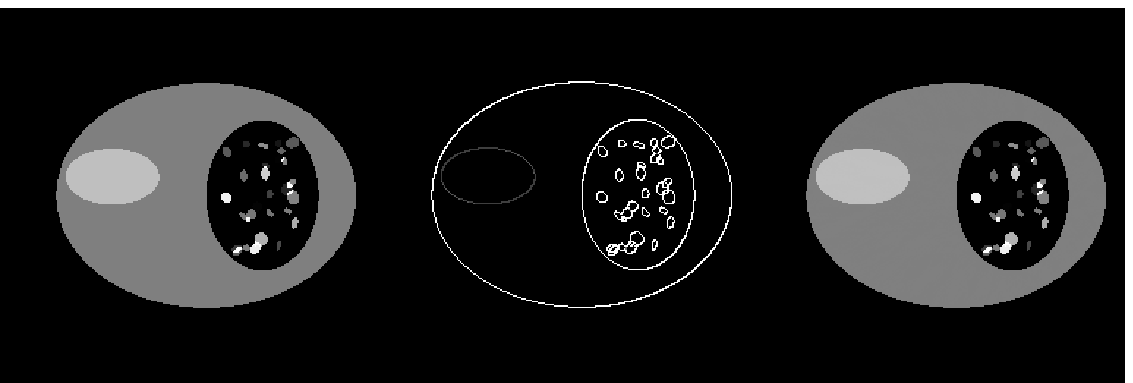}}
\end{minipage}
\begin{minipage}[b]{0.8\linewidth}
\centering
\centerline{\includegraphics[width=12cm,clip=TRUE]{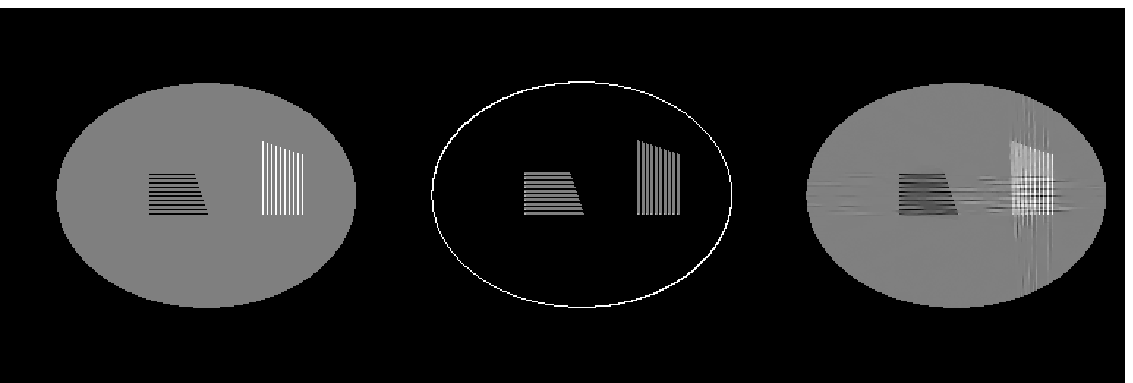}}
\end{minipage}
\caption{Images for the random ellipses (upper row),
the random spots (middle row), and lines (lower row) phantoms.
The true and gradient images of these phantoms are displayed 
in columns one and two, respectively. Images reconstructed from
20-view projections by use of the TV algorithm
are displayed in column three.
The gray scales for the images are [0.95, 1.15] for row one and 
[0.9, 1.1] for rows two and three.
\label{fig:otherPhantoms}}
\end{figure}

In an attempt to demonstrate
the wide applicability of the TV algorithm, we have also
applied it -- without changing any parameters in the algorithm --
to three additional phantoms, as shown in the first column
of Fig. \ref{fig:otherPhantoms}. 
%The results of these other tests are shown 
%in Fig. \ref{fig:otherPhantoms}.
%
The properties of these phantoms are as follows. The ``random ellipse''
phantom consists of 10 randomly selected ellipses on a uniform circular
background with a value of 1.0.  The values
of each of the ellipses was randomly selected in the range of [1.01, 1.10].
The ``random spots'' phantom is similar in that 30 randomly selected small
ellipses within the value range of [0.9, 1.1] are placed in an air cavity.
The background ellipse has a value of 1.0 and and additional ellipse
with a value of 1.05 is placed on the left of the phantom.  The spots and 
the air gap are meant to resemble, roughly, the lung. The ``lines'' phantom
consists of 2 groups of 10 lines at values of 0.9 and 1.1 on
a background ellipse of value 1.0.  As with the other phantoms,
the gradient image of the lines phantom has sparse structures
as shown in the second column
of Fig. \ref{fig:otherPhantoms}.
But the lines phantom is
designed in such a way as to provide a stiff challenge for the TV
algorithm. It is known for the FT-inversion problem
that certain regular structures in the image may be difficult to 
reconstruct by use of the FT-TV algorithm because of the small 
support of such images in Fourier space \cite{candes-robust}. 
We expect that such images also 
pose a challenge for the TV algorithm developed in this work.

Using these phantoms, we generated fan-beam projection data at 
20 views (uniformly distributed over $2 \pi$, specified by
Eq. (\ref{fewViewAngles})).
We show in column three of Fig. \ref{fig:otherPhantoms}
the TV reconstructions for
the random ellipses (upper row),
the random spots (middle row), and lines (lower row) phantoms.
The reconstructions for the random ellipses and random spots 
phantoms are visually indistinguishable from their corresponding
truth.  As expected the lines phantom proves to be
challenging. Although the reconstruction for the 
lines phantom does show some artifacts, it is still impressive. 
A quick glance at the EM and ART results in Fig. \ref{fig:fewView}  
will remind the reader how unstable image reconstruction 
is for this few-view scanning configuration.  

\subsection{Limited-angle problems}

An important application of the TV algorithm may be for reconstruction
problems where there are insufficient data in the corresponding continuous
case. For example, the scanning angle may be less than 180$^\circ$ plus the
fan angle in fan-beam CT,
or there may be gaps on the detector for each projection when
the data are known to be bad for certain detector bins.
For continuous functions of compact support it is well-known that
data in a scanning range of 180$^\circ$ 
plus the fan-angle is sufficient for stable image reconstruction
in fan-beam CT.
For the fan-beam configuration described above, 180$^\circ$
plus the fan angle is 209$^\circ$.  For scanning angular ranges less
than 209$^\circ$, the corresponding discrete linear system
is likely to be generally be ill-posed.

\begin{figure}[ht]

\begin{minipage}[b]{0.8\linewidth}
\centering
\centerline{\includegraphics[width=15cm,clip=TRUE]{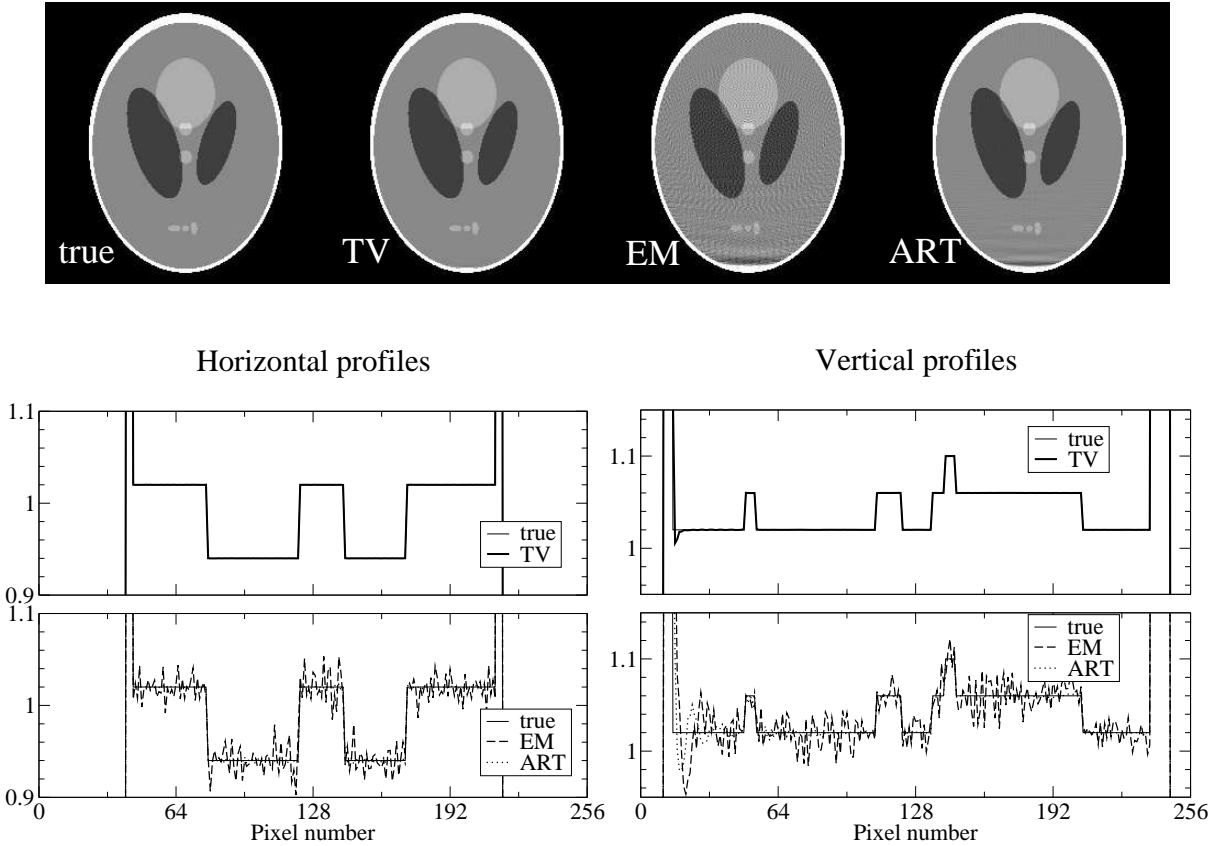}}
\hskip 1.0cm
\end{minipage}
\begin{minipage}[b]{0.48\linewidth}
\centering
\centerline{\includegraphics[width=8cm,clip=TRUE]{figs/piX.eps}}
\end{minipage}
\begin{minipage}[b]{0.48\linewidth}
\centering
\centerline{\includegraphics[width=8cm,clip=TRUE]{figs/piY.eps}}
\end{minipage}
\caption{
Upper row: The true image and images reconstructed by use of the 
TV, EM, and ART algorithms
from data over 180$^\circ$.
The display gray scale is [0.85,1.15]. 
Middle row: Image profiles along the centers
of the images in the horizontal  and 
vertical directions obtained with the TV algorithm (thick line).  
Lower row: Image profiles along the centers
of the images in the horizontal and 
vertical directions obtained with the EM
(dashed lines) and ART (dotted lines) algorithms.  
The corresponding true profiles are plotted as the
thin lines in the middle and lower rows. 
\label{fig:pi}}
\end{figure}

In the first limited-angle problem, we reduce the scanning 
angular range from 209$^\circ$ to 180$^\circ$ and
generate
projection data at 128 uniformly distributed views
from the Shepp-Logan phantom.
Again, the detector at each view has 512 bins.  For 
this scan, the number of non-zero data points is 
52,730 , which is even more than the number of non-zero 
pixels in the Shepp-Logan phantom itself.  

We display in the upper row of Fig. \ref{fig:pi} images reconstructed 
from this set of data by use of the TV, EM, and ART algorithms.
The profiles of these images along the central horizontal and
vertical rows are displayed in the middle and lower rows.
The number of iterations for each of the TV, EM, and
ART reconstructions is 1000. 
The images in row one of Fig. \ref{fig:pi} show that
the TV reconstruction is virtually
indistinguishable from the true phantom
and that the images obtained by use of the EM and ART 
algorithms are also reasonably accurate with only 
small distortion near the bottom of the images.
This distortion of the EM and ART images is understandable because
the 180$^\circ$ scan covered the top half of the phantom.  
The high iteration numbers were used for achieving convergence
in the bottom half of the image.  Additionally, the EM image 
shows a high frequency artifact not seen in the TV or ART images,
because the back-projector in each case is ray-driven, which
is known to yield such Moire patterns in EM images \cite{Basu:04}. 
But, as explained above, we are comparing
the reconstruction algorithms on their ability to solve the linear
system corresponding to the imaging model, and we therefore use the
ray-driven backprojection because it represents  
exactly the system-matrix adjoint.
% to be the exact transpose of the system matrix.

\begin{figure}[ht]

\begin{minipage}[b]{0.8\linewidth}
\centering
\centerline{\includegraphics[width=15cm,clip=TRUE]{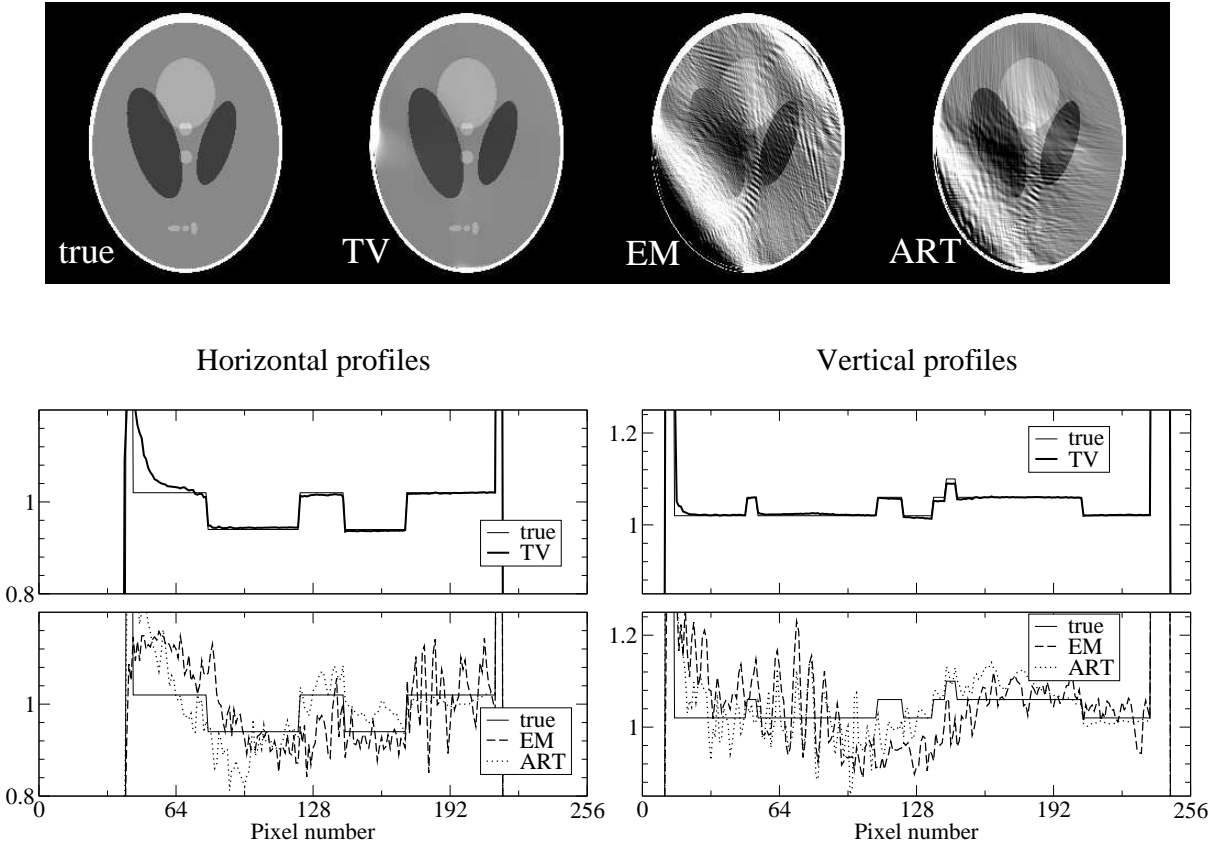}}
\hskip 1.0cm
\end{minipage}
\begin{minipage}[b]{0.48\linewidth}
\centering
\centerline{\includegraphics[width=8cm,clip=TRUE]{figs/piO2X.eps}}
\end{minipage}
\begin{minipage}[b]{0.48\linewidth}
\centering
\centerline{\includegraphics[width=8cm,clip=TRUE]{figs/piO2Y.eps}}
\end{minipage}
\caption{
Upper row: The true image and images reconstructed by use of the 
TV, EM, and ART algorithms
from data over 90$^\circ$.
The display gray scale is [0.85,1.15]. 
Middle row: Image profiles along the centers
of the images in the horizontal  and 
vertical directions obtained with the TV algorithm (thick line).  
Lower row: Image profiles along the centers
of the images in the horizontal and 
vertical directions obtained with the EM
(dashed lines) and ART (dotted lines) algorithms.  
The corresponding true profiles are plotted as the
thin lines in the middle and lower rows. 
\label{fig:piO2}}
\end{figure}

We explore further reduction in the scanning angle by taking
64 angular samples uniformly distributed over an 
angular range of only 90$^\circ$, covering the first
quadrant of the Shepp-Logan phantom in Fig. \ref{fig:piO2}.
We display in row one of Fig. \ref{fig:piO2} images reconstructed
by use of the TV, EM, and ART algorithms.
The number of iterations for the TV, EM, and ART reconstructions 
is 10,000. In this case, there were 26,420 non-zero projection
measurements, which would seem to be sufficient for 
the TV algorithm considering the sparseness of the phantom's image
gradient.  But the instability of the corresponding linear
system appears to be too strong for accurate
image reconstruction as can be seen in the reconstructions shown 
in the upper row of Fig. \ref{fig:piO2}. 
In the middle row of Fig. \ref{fig:piO2},
we show the profiles along central lines in 
the horizontal and vertical directions 
of the TV image. The corresponding true profiles are also
displayed as the thin lines. 
The TV image 
contains a deviation from the true phantom
on the left-hand edge, which is evident
in the shown horizontal profile.  
On the other hand, the EM and ART reconstructions
are highly distorted. We have studied 
in row three of Fig. \ref{fig:piO2}
the profiles along central lines in 
the horizontal and vertical directions 
of the EM and ART
images. Distortions in these images are clearly 
shown in these profile plots. We  
have also studied the image error 
as a function of iteration number in an effort to determine
whether or not the TV algorithm will converge to the true image.
For the previous cases the image error was tending to zero, but
for this 90$^\circ$ scan the image error appears to converge
to a small but finite positive number. 
The system matrix corresponding to the 90$^\circ$ scan 
appears to violate somewhat the ERP.

\subsection{Projection data with ``bad'' detector bins}
\label{sec:badBins}

\begin{figure}[ht]

\begin{minipage}[b]{\linewidth}
\centering
\centerline{\includegraphics[width=7cm,clip=TRUE]{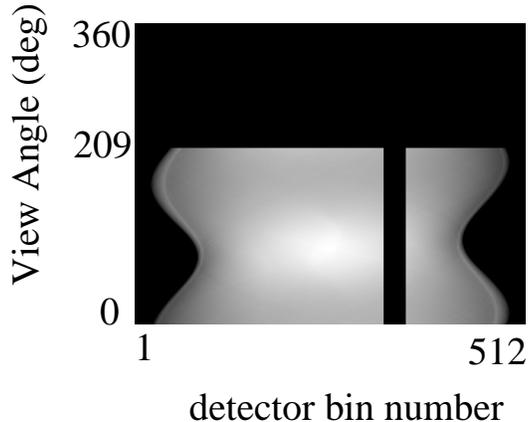}}
\end{minipage}
\caption{
Intensity plot of the ``bad bins'' projection data function. The angular
range covers 209$^\circ$, which is the short-scan angle
for the current fan-beam configuration. However, data at 30 of the 
512 detector bins are missing.
\label{fig:badBinsSino}}
\end{figure}

Another reconstruction problem of practical interest is how to handle
the situation where data from a set of bins on the detector are 
corrupted.  Such a problem could occur if there is
a hardware failure or if the photon count is very low so that signal
noise dominates.  For fan-beam CT,
if a full scan is performed over 360$^\circ$, one may
fill the gaps in the detector bins by using redundant data 
at conjugate views. For a short-scan, however, this  approach may
not be possible. Specifically, consider projection data displayed in
Fig.~\ref{fig:badBinsSino}; the angular range scanned is the minimum
for exact reconstruction, namely, 180$^\circ$ plus the fan angle,
which in this case is a total of $209^\circ$. The projection data
at each view, however, has a gap.  Because the scanning angle is over
the minimum range, there may not be redundant information to fill in the
gap left by the ``bad'' detector bins.  Direct application of analytic
algorithms such as fan-beam FBP will yield
conspicuous artifacts, as the implicit assumption is that the missing
values are zero, which is highly inconsistent with the rest of the data
function.

We apply the TV algorithm to reconstructing images from data shown
in Fig. \ref{fig:badBinsSino}, which are generated at 150 views 
uniformly distributed over 209$^\circ$.
The detector at each view 
contains 512 bins, of which the data of 30 bins have been discarded
as shown in Fig. \ref{fig:badBinsSino}.  Again, in this case there may 
be enough data to determine the image, because the number 
of non-zero projection measurements is 58,430.  
The question is whether or not the corresponding
linear system is stable
enough that the solution can be found.

\begin{figure}[ht]

\begin{minipage}[b]{0.8\linewidth}
\centering
\centerline{\includegraphics[width=15cm,clip=TRUE]{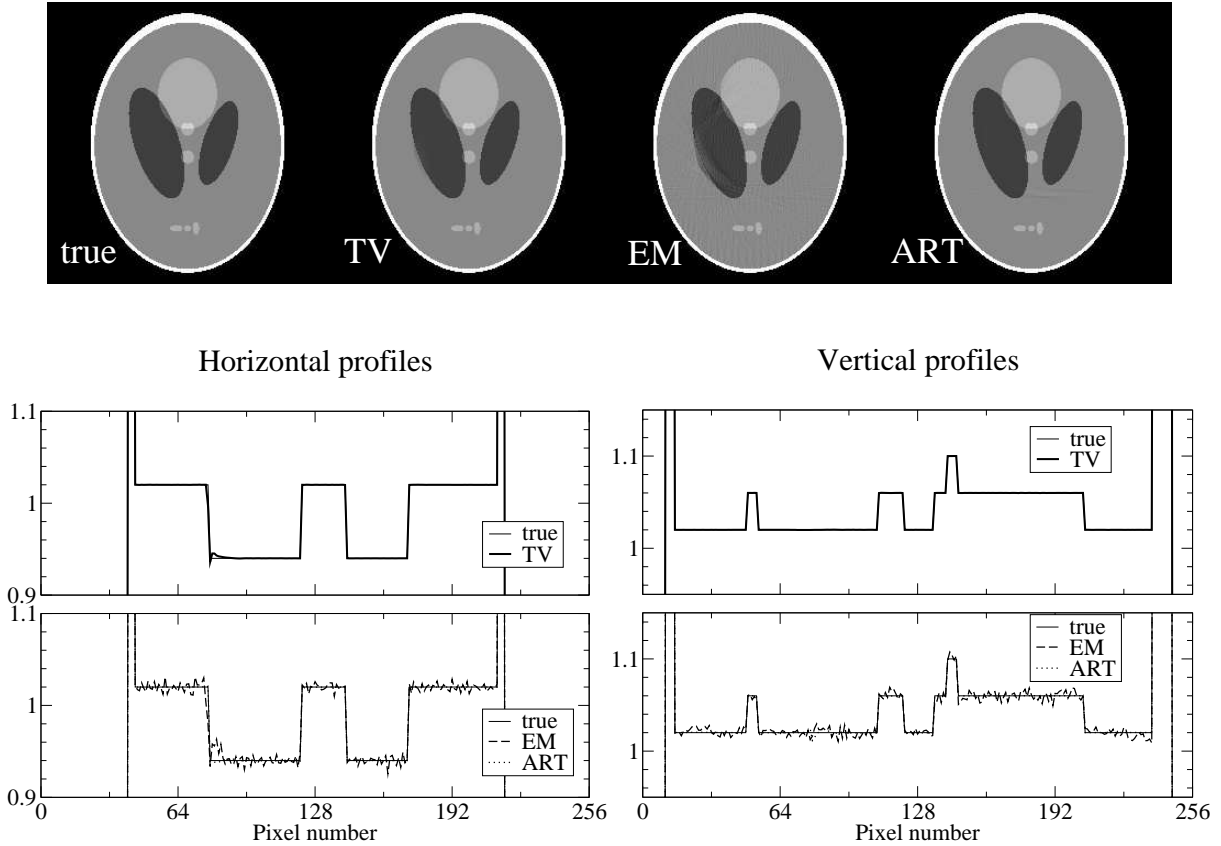}}
\hskip 1.0cm
\end{minipage}
\begin{minipage}[b]{0.48\linewidth}
\centering
\centerline{\includegraphics[width=8cm,clip=TRUE]{figs/badBinsX.eps}}
\end{minipage}
\begin{minipage}[b]{0.48\linewidth}
\centering
\centerline{\includegraphics[width=8cm,clip=TRUE]{figs/badBinsY.eps}}
\end{minipage}
\caption{
Upper row: The true image and images reconstructed by use of the 
TV, EM, and ART algorithms
from data containing bad detector bins.
The display gray scale is [0.85,1.15]. 
Middle row: Image profiles along the centers
of the images in the horizontal  and 
vertical directions obtained with the TV algorithm (thick line).  
Lower row: Image profiles along the centers
of the images in the horizontal and 
vertical directions obtained with the EM
(dashed lines) and ART (dotted lines) algorithms.  
The corresponding true profiles are plotted as the
thin lines in the middle and lower rows. 
\label{fig:badBins}}
\end{figure}

We display in Fig. \ref{fig:badBins} images 
reconstructed by use of the TV, EM, and ART
algorithms. 
Once again, the TV image is visually indistinguishable
from the true image, and both EM and ART algorithms yield
in this case quite accurate images. In this study, 
the TV algorithm appears to be more robust than the EM and 
ART algorithms, because the TV image 
is obtained with only a 100 
iterations while both the EM and ART algorithms
required 10000 iterations to achieve the image accuracy shown
in Fig. \ref{fig:badBins}.
We note that the previous FT-TV algorithm cannot address the bad bins
problem directly even in the parallel-beam case, because it is not
possible to perform the FT of the detector data at each view
when there is a gap.

\subsection{Few-view projection data with bad bins}

The previously discussed insufficient data problems can
be combined. For example, we consider the few-view problem discussed
in Sec. \ref{sec:fewView} with each projection view containing bad bins,
as in the previous section.  For this experiment we take projections at 20 views
uniformly covering the short-scan angular 
range with the same detector gap as shown
in Fig. \ref{fig:badBinsSino}.  Thus the difference between this study and
the one of Sec. \ref{sec:badBins} is that the angular spacing between
projections here is roughly 10$^\circ$ instead of the 1.4$^\circ$ spacing
in the previous section. The few-view-projection data
are sparse, and only 7735 measured data points are non-zero.

\begin{figure}[ht]

\begin{minipage}[b]{0.8\linewidth}
\centering
\centerline{\includegraphics[width=15cm,clip=TRUE]{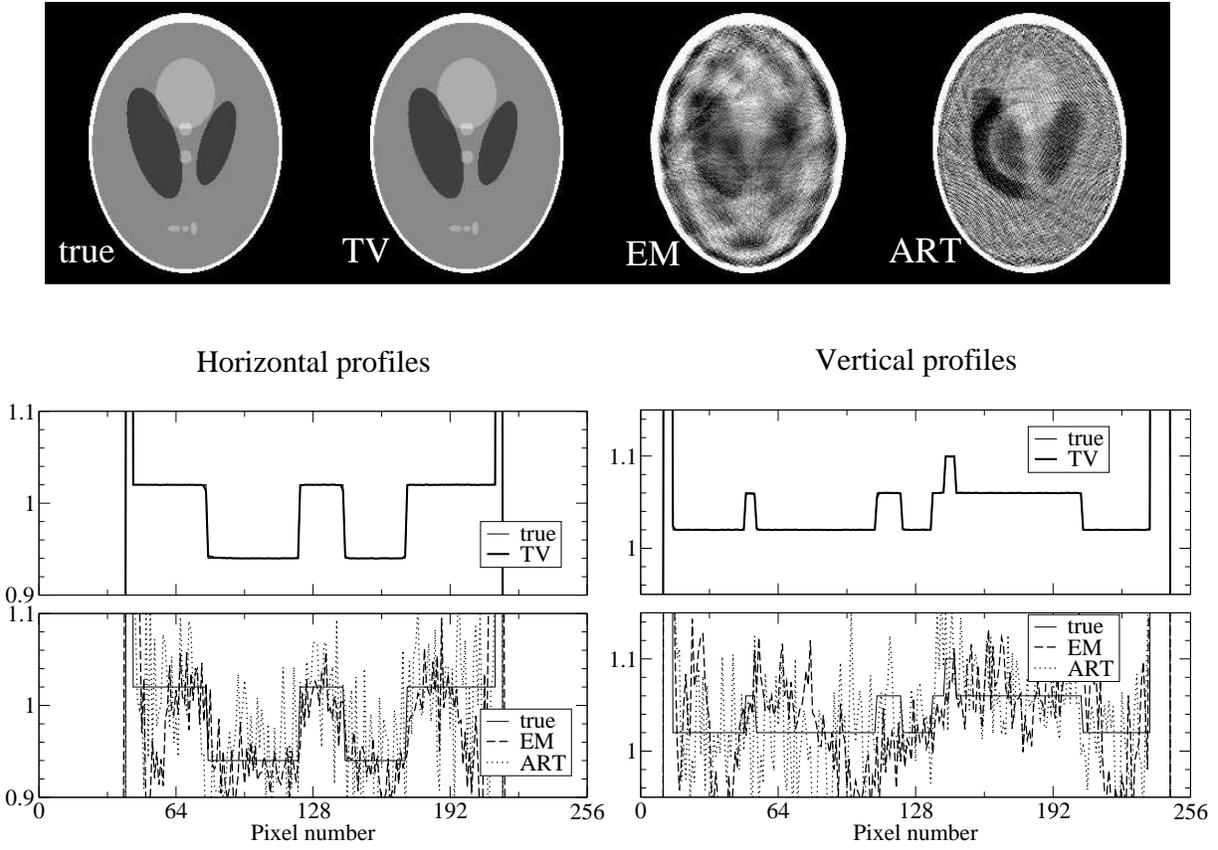}}
\hskip 1.0cm
\end{minipage}
\begin{minipage}[b]{0.48\linewidth}
\centering
\centerline{\includegraphics[width=8cm,clip=TRUE]{figs/bbFewViewX.eps}}
\end{minipage}
\begin{minipage}[b]{0.48\linewidth}
\centering
\centerline{\includegraphics[width=8cm,clip=TRUE]{figs/bbFewViewY.eps}}
\end{minipage}
\caption{
Upper row: The true image and images reconstructed by use of the 
TV, EM, and ART algorithms
from 20-view data containing bad detector bins.
The display gray scale is [0.85,1.15]. 
Middle row: Image profiles along the centers
of the images in the horizontal  and 
vertical directions obtained with the TV algorithm (thick line).  
Lower row: Image profiles along the centers
of the images in the horizontal and 
vertical directions obtained with the EM
(dashed lines) and ART (dotted lines) algorithms.  
The corresponding true profiles are plotted as the
thin lines in the middle and lower rows. 
\label{fig:fewViewBB}}
\end{figure}

We show in Fig. \ref{fig:fewViewBB} images 
reconstructed by use of the TV, EM, and ART
algorithms.
The TV image is once again visually
indistinguishable from the true phantom.
Thus, it appears that the system matrix
corresponding to this scanning configuration fulfills the ERP.
The EM and ART reconstructions show similar artifacts as 
were seen in the few-view results shown
in Sec. \ref{sec:fewView}.  In addition, there appears to be additional
artifacts from the missing detector bins.

The proposed TV algorithm can address a host of other sparse data 
problems. The key points for the success of the algorithm 
-- under ideal conditions described above --
are that the support of the data function be at least twice the support
of the gradient of the true image and that the corresponding linear
system is not too ill-conditioned as was seen for 
the 90$^\circ$-scan case.

\section{Numerical Results: complicating factors}
\label{sec:compfac}

The results of the previous section assumed the ideal situation of
perfect consistency among the measured projection rays and a sufficiently
sparse gradient image.  We show below how
the TV, EM, and ART algorithms compare when these conditions are
not strictly held,
by adding a varying background, to violate gradient 
sparseness, or by adding signal noise, to violate 
data consistency.

\subsection{Violation of gradient sparseness}

In many applications the gradient images may be sparse only
in an approximate sense.  Even though it is a good approximation to assume that
images will be constant over many regions,
there will also be situations in which the images will have
some level of variation within the regions.
An important question is whether or not a low amplitude
violation of gradient sparseness leads
to only small deviations in images reconstructed by
use of the TV algorithm. We investigate this issue 
by repeating the few-view and bad-bin studies described 
in Secs. \ref{sec:fewView} and \ref{sec:badBins}, but
adding a wavy background to the Shepp-Logan 
phantom. 

\begin{figure}[ht]

\begin{minipage}[b]{0.8\linewidth}
\centering
\centerline{\includegraphics[width=15cm,clip=TRUE]{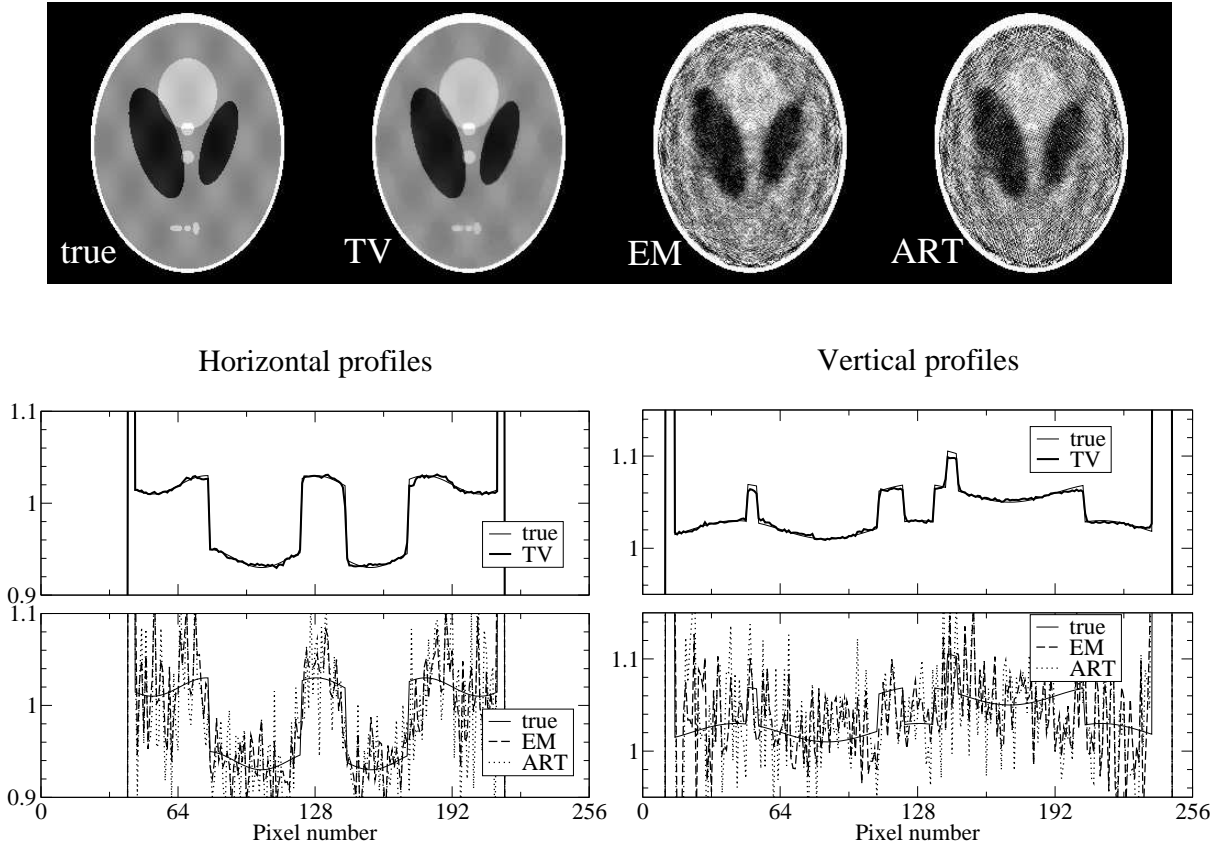}}
\hskip 1.0cm
\end{minipage}
\begin{minipage}[b]{0.48\linewidth}
\centering
\centerline{\includegraphics[width=8cm,clip=TRUE]{figs/fewViewWaveyX.eps}}
\end{minipage}
\begin{minipage}[b]{0.48\linewidth}
\centering
\centerline{\includegraphics[width=8cm,clip=TRUE]{figs/fewViewWaveyY.eps}}
\end{minipage}
\caption{
Upper row: The true image with a wavy background
and images reconstructed by use of the 
TV, EM, and ART algorithms
from 20-view data.
The display gray scale is [0.85,1.15]. 
Middle row: Image profiles along the centers
of the images in the horizontal  and 
vertical directions obtained with the TV algorithm (thick line).  
Lower row: Image profiles along the centers
of the images in the horizontal and 
vertical directions obtained with the EM
(dashed lines) and ART (dotted lines) algorithms.  
The corresponding true profiles are plotted as the
thin lines in the middle and lower rows. 
\label{fig:fewViewWavey}}
\end{figure}

Using the Shepp-Logan phantom with a wavy background in  
Fig. \ref{fig:fewViewWavey}, we generated projection
data at 20 views specified by Eq. (\ref{fewViewAngles}).  
The amplitude of the wavy background
is 1\% of the gray matter attenuation coefficient.  Any negative values
in the phantom are thresholded to zero, so as to allow the applicability
of the EM algorithm.  With the wavy background the number of non-zero
pixels in the gradient image jumps to 51,958 , but the majority of these
non-zero values are small compared to the gradients at the boundaries of
the different tissues.  As was the case with the previous few-view study,
the number of measurements is 10,240, which is less than twice the number
of non-zero pixels in the gradient image, violating the sparseness
condition.

In Fig. \ref{fig:fewViewWavey}, we show the 
images reconstructed by use of the TV, EM, and
ART algorithms from the 20-view data.
The iteration numbers for obtaining these results were 200, 1000,
and 500 for the TV, EM, and ART algorithms, respectively.
The images in Fig. \ref{fig:fewViewWavey} indicate 
that the TV reconstruction is visually almost indistinguishable
from the true image and that the EM and ART algorithms 
have difficulty with this data set.  
Upon further inspection of the image profiles, it can be
seen that the TV algorithm
does not yield an exact reconstruction. The small violation,
however, of the
gradient image sparseness does not appear to lead to large 
errors in the reconstructed image.
We point out once again that this example does not constitute a 
mathematical proof, but it
is suggestive of the conclusion
that small violations in the gradient sparseness 
yields only small errors in the reconstructed image.

\begin{figure}[ht]

\begin{minipage}[b]{0.8\linewidth}
\centering
\centerline{\includegraphics[width=15cm,clip=TRUE]{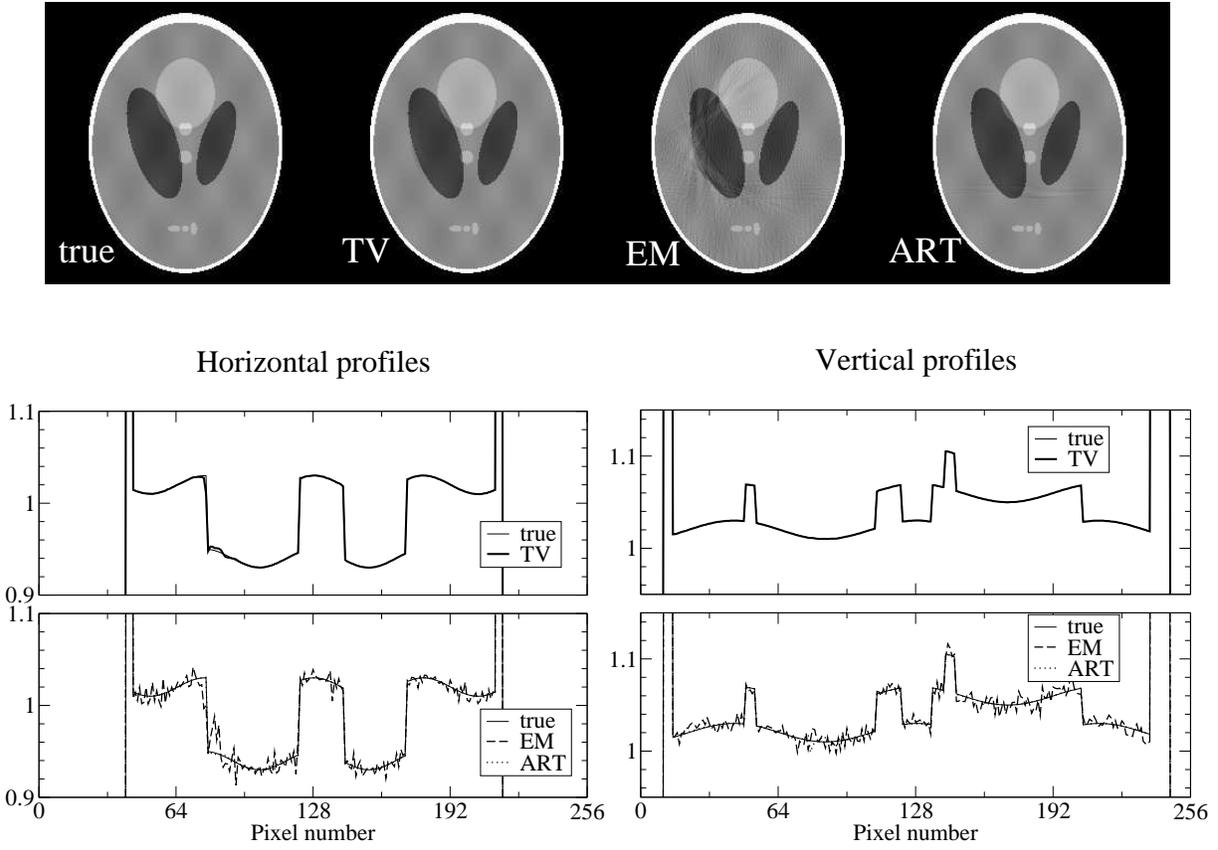}}
\hskip 1.0cm
\end{minipage}
\begin{minipage}[b]{0.48\linewidth}
\centering
\centerline{\includegraphics[width=8cm,clip=TRUE]{figs/badBinsWaveyX.eps}}
\end{minipage}
\begin{minipage}[b]{0.48\linewidth}
\centering
\centerline{\includegraphics[width=8cm,clip=TRUE]{figs/badBinsWaveyY.eps}}
\end{minipage}
\caption{
Upper row: The true image with a wavy background
and images reconstructed by use of the 
TV, EM, and ART algorithms
from bad detector bin data.
The display gray scale is [0.85,1.15]. 
Middle row: Image profiles along the centers
of the images in the horizontal  and 
vertical directions obtained with the TV algorithm (thick line).  
Lower row: Image profiles along the centers
of the images in the horizontal and 
vertical directions obtained with the EM
(dashed lines) and ART (dotted lines) algorithms.  
The corresponding true profiles are plotted as the
thin lines in the middle and lower rows. 
\label{fig:badBinsWavey}}
\end{figure}

We also reexamined image reconstruction from data containing
bad-bins of Sec. \ref{sec:badBins} with the $1\%$ low amplitude
wavy background added to the original image.  
In this case, the number of projection data is 58,430, which
is not twice the number of non-zero pixels in the image but it is 
a comparable number. 
We display in Fig. \ref{fig:badBinsWavey} images 
reconstructed by use of the TV, EM, and ART
algorithms.  It can be observed that 
the TV image is visually indistinguishable
from the true image.
We also note that, as before, the ART and EM reconstructions
are close to the original image in this case. 
The number of iterations for the TV algorithm is 100, 
which is much less than the 10,000 iterations used
for both EM and ART algorithms.

\subsection{Reconstruction from noisy data}

Another omni-present physical factor that contributes to data inconsistency
is signal noise in the projection measurements.  
It is of practical significance
to evaluate the performance of the TV algorithm in the presence of data noise.
While a thorough evaluation of the noise properties of the TV
algorithm is beyond the scope of this work, 
we present preliminary results indicating
that the TV algorithm appears to be effective on sparse data problems even
when the data contain inconsistencies due to signal noise.
For the noise studies, we again take the few-view and bad-bin
cases in Secs. \ref{sec:fewView} and \ref{sec:badBins}.
In each case, Gaussian noise is introduced in the projection 
data at the level of 0.1$\%$ of the ideal measurement values.

The total variation algorithm has an interesting and practical
feature with respect to data sets that contain inconsistencies.
Even though TV minimization is part
of a larger algorithm that implements the program in Eq. (\ref{optimization}),
the gradient descent phase happens to also regularize the image.
This feature of the TV algorithm is particularly useful in the 
present case of data inconsistencies. As a result, for this section
we present two images from the TV algorithm: $f^{(TV-GRAD)}[n,N_\text{grad}]$,
the image after the completion of the gradient descent phase, and
$f^{(TV-POS)}[n]$, the image after the completion of the POCS phase.
The former image is labeled TV1 and the latter TV2. The TV1 image
is seen to be a regularized version of the TV2 image.

\begin{figure}[ht]

\begin{minipage}[b]{0.8\linewidth}
\centering
\centerline{\includegraphics[width=15cm,clip=TRUE]{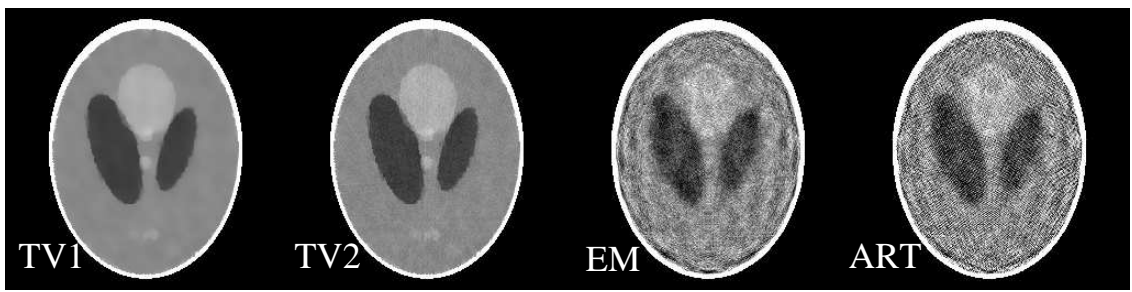}}
\end{minipage}
\caption{
Images reconstructed from 20-view noisy data
by use of the TV algorithm after the gradient descent phase 
(TV1) and after the projection phase (TV2) and by use of 
the EM and ART algorithms.
\label{fig:fewViewNoisy}}
\end{figure}

For the few-view study, we show in 
Fig. \ref{fig:fewViewNoisy} images reconstructed by use of 
the TV (labeled TV1 and TV2), EM,
and ART algorithms. The iteration numbers
for the TV, EM and ART images are 200, 200, and 100, respectively.  
In the studies with consistent data above, 
the differences between the TV1 and TV2 images were numerically
negligible. With inconsistencies resulting from data noise, 
however, there is a marked difference.  The image
$f^{(TV-GRAD)}[n,N_\text{grad}]$ after the gradient descent phase 
is clearly a regularized version of the image 
$f^{(TV-POS)}[n]$
obtained after the data projection and positivity constraint.  
Depending on the task, either image may prove
useful for a particular imaging application. 
For the few-view study, both
images $f^{(TV-GRAD)}[n,N_\text{grad}]$ and
$f^{(TV-POS)}[n]$
obtained with the TV algorithm appear to have less artifacts than
the EM and ART reconstructions  
in Fig. \ref{fig:fewViewNoisy}.
We point out again that no explicit regularization
is performed with EM or ART
in the studies here and below
aside from the fact that we
truncate the iteration numbers at 200 and 100
in the EM and ART algorithms, respectively. 

\begin{figure}[ht]

\begin{minipage}[b]{0.8\linewidth}
\centering
\centerline{\includegraphics[width=15cm,clip=TRUE]{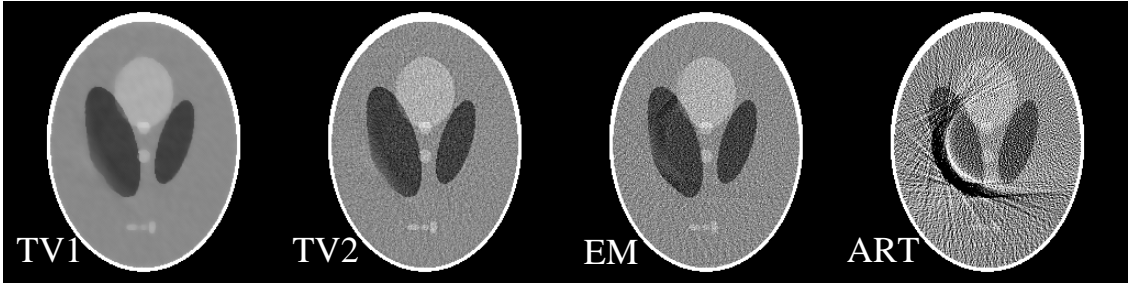}}
\end{minipage}
\caption{
Images reconstructed from bad-bin noisy data
by use of the TV algorithm after the gradient descent phase 
(TV1) and after the projection phase (TV2) and by use of 
the EM and ART algorithms.
\label{fig:badBinsNoisy}}
\end{figure}

For the bad bin case, we generated noisy data by adding
Gaussian noise, at the level of 0.1\% of the individual true
data values, to the noiseless 
data described in Sec. \ref{sec:badBins}. In  
Fig. \ref{fig:badBinsNoisy}, we show images reconstructed by use of
the TV algorithm (TV1) and (TV2), the EM,
and the ART algorithm. The iteration numbers
for the TV, EM, and ART images are 200, 200, and 100, respectively.  
Again, we show two TV images in Fig. \ref{fig:fewViewNoisy}:
TV1 and TV2.
The results of this study
suggests that the TV and EM algorithms can still effectively
correct for the effect of the missing detector bins.  
The ART algorithm, which showed very mild streaking 
in Fig. \ref{fig:badBins} under the ideal condition,
displays significant streaking due to the combination 
of signal noise and bad detector bins.

\section{Discussion and Conclusion}

In this article, we have developed a TV algorithm
for accurate image reconstruction in divergent-beam 
CT under a number of imperfect sampling situations.  
We have evaluated and demonstrated the performance
of the TV algorithm in addressing a number of challenging
reconstruction problems, including the 
few-view, limited-angle, and bad-bin problems.
As results in these numerical studies indicate, 
the proposed TV algorithm can yield accurate reconstructions
in these difficult cases, which are of practical
significance. The effectiveness of the TV algorithm
relies on the fact that the object being imaged has a relatively
sparse gradient image.  It should be pointed out that
we did not provide a theoretical proof of the ERP conjecture
for the various scanning configurations studied here; however,
we speculate based on the numerical examples
that this principle 
may apply to many insufficient data problems in divergent-beam CT.
In future work we will compare the TV algorithm with state-of-the-art
implementations of EM and ART, and compare with algorithms that
have been designed to handle few-view \cite{Kudo:02,Wu:03,KoleI:03} and
limited angular range \cite{PerssonLimited:01,Bresler:98,Rantala:06} problems.
The TV algorithm described
above applies equally to cone-beam CT, even though our examples 
were limited to fan-beam CT.  The TV algorithm may also prove
useful for other tomographic imaging modalities.

There are numerous aspects of the TV algorithm that may make it relevant
and useful for medical and industrial CT imaging.  The assumption 
of a sparse gradient image is quite reasonable
for many object functions in medical and industrial
applications, because often sought-after quantities such as
x-ray attenuation coefficient are relatively
constant over extended areas or volumes.
We showed example reconstructions from data containing
two of the most likely imperfections.  First, one can expect
that the sparseness of the image gradient will hold only
approximately, and second, there will always be some level of
inconsistency among the projection data due to signal noise.
Our numerical studies with respect to these complicating factors 
appear to show that the TV algorithm can effectively reconstruct
quantitatively accurate images from imperfectly sampled data.  
We are currently investigating the application
of the TV algorithm to 3D cone-beam CT where there are a host
of imperfect sampling situations that have practical significance.
We will also investigate and develop refinements to the TV 
algorithm that optimize its performance.
% in order to decrease the number of iterations.

\begin{acknowledgments}
EYS was supported by National Institutes of Health grant
K01 EB003913.  CMK was supported in part by ACS-IL 05-18.
This work was also supported in part by
NIH grants R01 EB00225 and
R01 EB02765. Its contents are
solely the responsibility of the authors and do not necessarily
represent the official views of the National Institutes of Health.
\end{acknowledgments}

\bibliographystyle{apsrev}
%\bibliography{strings,refs,manuals}
\bibliography{tv}

\end{document}